\documentclass[aps,pre,twocolumn,superscriptaddress,showpacs,showkeys ]{revtex4-1}

\usepackage{graphicx}                   
\usepackage{hyperref}                   
\usepackage{amsmath,amssymb}            
\usepackage{epstopdf}
\usepackage{subcaption}					

\usepackage{color}
\definecolor{orange}{rgb}{1,0.5,0}
\definecolor{brown}{rgb}{0.65, 0.16, 0.16}
\definecolor{phlox}{rgb}{0.87, 0.0, 1.0}

\graphicspath{{figs/}}					

\begin{document}

\title{Coupling of $c=-2$ and $c=\frac{1}{2}$ and $c=0$ conformal field theories:\\
	 The geometrical point of view}

\author{M. N. Najafi}
\affiliation{Department of Physics, University of Mohaghegh Ardabili, P.O. Box 179, Ardabil, Iran}
\email{morteza.nattagh@gmail.com}

\begin{abstract}
The coupling of the $c=-2$, $c=\frac{1}{2}$ and $c=0$ conformal field theories are numerically considered in this paper. As the prototypes of the couplings, $(c_1=-2)\oplus (c_2=0)$ and $(c_1=-2)\oplus (c_2=\frac{1}{2})$, we consider the BTW model on the 2D square critical site-percolation and the BTW model on Ising-correlated percolation lattices respectively. Some geometrical techniques are used to characterize the presumable conformal symmetry of the resultant systems. Using the Schramm-Loewrner evolution (SLE), we find that the algebra of the central charges of the coupled models is closed, namely the former case results to two-dimensional self-avoiding walk (SAW) fixed point corresponding to $c=0$ conformal field theory, whereas the latter one results to the two-dimensional critical Ising fixed point corresponding to the $c=\frac{1}{2}$ conformal field theory.
\end{abstract}

\pacs{05., 05.20.-y, 05.10.Ln, 05.65.+b, 05.45.Df}
\keywords{Bak-Tang-Wiesenfeld model, Ising correlations, percolation lattice, self-avoiding walk}

\maketitle

\section{Introduction}
The interplay between the critical models is a challenging problem in the context of statistical field theory and integrable models. Examples of coupling two critical models are critical dynamics of sandpiles on various lattices~\cite{Huynh2011Abelian}, on the quenched substrate generated by kinetic self-avoiding trails~\cite{Karmakar2005sandpiles}, on the two-dimensional percolation lattice~\cite{Najafi2016Bak,Najafi2013water} and the Ising ferromagnet on percolation lattice~\cite{Najafi2016Monte}. In these examples a critical dynamical model has been defined on a (metric) system in which the diluteness pattern is modeled by another critical model. This coupling of two critical models may be seen as a deformation of a fixed point (the original critical models) which is perturbed by means of a scaling field corresponding to the other critical model~\cite{Mussardo2010Statistical,Eguchi1989Deform,Cardy1990Form}. In this view, the Zamolodchikov's $c$-theorem helps to identify the structure of these fixed points provided that the perturbing operator of the critical action is known. More precisely the Zamolodchikov's $c$-theorem which applies for the \textit{off-critical} conformal field theories is able to yield the change of the corresponding central charge $\Delta c\equiv c_{\text{IR}}-c_{\text{UV}}$ in which $c_{\text{IR}}$ and $c_{\text{UV}}$ are the central charges of the conformal field theories (CFT) of the IR and UV fixed points. This approach minimally needs the knowledge of the perturbing scaling field (or fields) by which the $c_{\text{UV}}$ CFT model becomes unstable towards the IR fixed point. For example perturbing a minimal model with central charge $c(p)\equiv 1-\frac{6}{p(p+1)}$ by means of the scaling primary field $\Phi_{13}$ results to a change of the central charge to $c_{\text{IR}}=1-\frac{6}{(p+1)(p+2)}$ which corresponds to the replacement $p\rightarrow p+1$~\cite{Mussardo2010Statistical}. Although this approach is well developed and much theoretical investigations have been done on the problem, it is not much practical for the general less-known models whose operator content are not known. Also this approach is not much understood for the logarithmic CFTs (LCFTs), e.g. the $c=-2$ and $c=0$ LCFTs. Therefore one should seek for a more direct method to find the presumable emergent CFT. \\
Thanks to the statistical techniques which has the ability to characterize the model in both (IR and UV) limits, the above mentioned problem can fruitfully be addressed. This can be achieved using the Schramm-Loewner evolution (SLE) technology by which the conformal symmetry of any 2D critical system can be addressed~\cite{Cardy}. This theory classifies the 2D critical models by a single parameter, named as the diffusivity parameter ($\kappa$) which is related directly to the central charge of the corresponding CFTs~\cite{BauBer}. Therefore, although determining the operator content of the model is not likely by the statistical analysis, the determination of the model existing in the IR limit as well as the UV limit is directly possible.\\
In this paper we consider the coupling of the $c=-2$, $c=0$ and $c=\frac{1}{2}$ CFTs. As the prototypes of these models, we consider respectively the 2D BTW model, the uncorrelated site percolation model and the 2D critical Ising model~\cite{francesco2012conformal}. The method of couplings is by implementing the dynamical critical (BTW) model on the critical lattices whose diluteness pattern is modeled by the critical percolation and the critical Ising models, to be described in the next section. The coupled models are studied via the SLE theory and some other tests. In coupling of the $c_1=-2$ and $c_2=\frac{1}{2}$ CFTs, we argue that the resulting system share some properties with the self-avoiding walk (SAW) which belongs to the $c=0$ CFT class. To this end, along with some statistical analysis such as the fractal dimension of the critical loops, we use the SLE technique for the open curves. We use the direct SLE mapping, as well as the left passage probability (LPP)~\cite{NajafiPRE1} and winding angel (WA)~\cite{Boffetta} tests to extract the diffusivity parameter $\kappa$. The obtained amount of $\kappa$ is compatible with the fractal dimension of loops (which are defined in the text).\\
In the other part of the paper we show numerically that the coupling of the $c_1=-2$ and $c_2=0$ CFTs is compatible with the $c=\frac{1}{2}$ CFT class. This result is precisely checked by the direct SLE mapping technique as well as the fractal dimension of the loops.\\
The paper has been organized as follows: In the next section we describe the method of coupling of the critical models. In the SEC~\ref{num} we describe the numerical methods and also the SLE theory. The SECs~\ref{sum1} and~\ref{sum2} have been devoted to the problem of $c=-2\oplus c=\frac{1}{2}$ and $c=-2\oplus c=0$ respectively.

\section{The Model}
\label{sec:model}

By coupling, two CFTs with say $c_1$ and $c_2$ central charges are indirectly added in such a way that their interplay cause a new model with a presumable conformal symmetry with the central charge $c\neq c_1+c_2$ (we note that $c=c_1+c_2$ occurs when the CFTs are trivially summed). We abbreviate this by $c_1\oplus c_2\equiv c$. For example consider $c_1=-2$ and $c_2=0$. The former refers to a LCFT whose most famous representative is the 2D BTW model and the latter also refers to a LCFT whose well-known prototype is 2D critical percolation. For the coupling of these two CFTs we can directly implement a $c_1=-2$ model on a lattice model which is described by the $c_2=0$ CFT. More precisely we simulate the BTW on the 2D critical site-diluted percolation lattice and extract the key parameters of the resultant system which determines its presumable universality class. We will present some evidences that it has conformal symmetry with the CFT central charge $c=\frac{1}{2}$, i.e. the 2D critical Ising universality class, which is in agreement with the previous result $-2\oplus 0\equiv \frac{1}{2}$~\cite{Najafi2016Bak}.\\
The other possible coupling is for the $c_1=-2$ and the $c_2=\frac{1}{2}$ CFTs, i.e.  $-2\oplus \frac{1}{2}$ whose motivation is to test the closeness of the algebra of the central charges. Realizing this problem is not as simple as the previous case, since the sand grains of the BTW model (as the $c_1=-2$ CFT) should interact with, e.g. the spins of the Ising model as the most famous representative of the $c_2=\frac{1}{2}$ CFT class. One way to construct such a coupling is to consider the BTW model as the dynamical model and the Ising model as the quenched random host system whose definition is presented in the remaining of this section. In fact we introduce a unified method to define the coupling $-2\oplus 0$ and $-2\oplus \frac{1}{2}$ both of which have the $c_1=-2$ CFT as the dynamical model.\\
Before going into the details of modeling the diluteness pattern of the metric space, let us define the BTW sandpile model and its definition on a general site-dilute square lattice. Sandpile models have been introduced by Bak et al~\cite{Bak1987Self} as an example for a class of models that show self-organized criticality. These models show critical behaviors, without tuning external parameters such as temperature. The Abelian structure of this model was first discovered by Dhar and named as the ASM~\cite{Dhar} and has vastly been studied in two as well as three dimensions~\cite{dashti2015statistical,Lubeck1997BTW}. Despite its simplicity, the ASM has various interesting features and many different analytical and numerical works have been performed on this model, for example, different height and cluster probabilities~\cite{Majumdar1} , its connection with spanning trees~\cite{Majumdar2}, ghost models~\cite{Mahieu}, q-state Potts model~\cite{Saleur,Coniglio}, etc. For a good review, refer to~\cite{Dhar2006Theoretical}. In the 2D regular lattice it is defined by attributing to each site some integer number $ \left\lbrace h_i\right\rbrace_{i=1}^N $ which is named as its height (or energy or the number of sand grains) with the restriction $1\le h_i\le 4\equiv h_c$ in which $h_c$ is some threshold and $N$ is the total number of sites. At the initial stage, one can set the heights at random. At each time step a grain is added to a randomly chosen site $i$, i.e. $ h_i \to h_i+1 $. If the height of this site exceeds $h_c$, a toppling occurs according to which $ h_i\to h_i+\Delta_{i,j} $ in which $\Delta_{i,j}=-4$ if $i=j$, $\Delta_{i,j}=1$ if $i$ and $j$ are neighbors and zero otherwise. A toppling can cause the nearest-neighbor sites to become unstable (have the height higher than $h_c$) and topple in their own turn and so on, until reaching the state in which all of the lattice sites become stable. This overall process is called an avalanche. The model is conservative and the energy is dissipated only from the boundary sites. After this relaxation process, another site is chosen and the process continues. There are two class of configurations: transient and recurrent. The former configurations take place once in an overall process, whereas the latter configurations have the chance of taking place more times. In the recurrent states, the statistical observables become constant with some statistical fluctuations. \\
For the cases in which the metric space is not perfect and regular the definition is trivially generalized. In this case the site can have one of two (quenched) states: active $s=1$ (into which the sand grains can enter) or inactive $s=0$ (which do not allow the sand grains to pass through them). Let us denote the effective local coordination number by $z_i\equiv\sum_{\delta}s_{i+\delta}$ (the number of active neighbors of an active (say $i$th) site), in which $\delta$ runs over all of the neighbors of the $i$th site. In this case the rule of a toppling at the site $i$ ($h_i>h_c$) is $ h_i\to h_i+\Delta_{i,j} $ in which $\Delta_{i,j}=-z_i$ if $i=j$, $\Delta_{i,j}=s_j$ if $i$ and $j$ are neighbors and zero otherwise. The same concepts (e.g. avalanche, steady states, transient and recurrent configurations) also hold for this case. \\
Now let us fix the lattice model as the CFT partner of the $c_1=-2$ CFT, which is done by identifying the model to set the quenched $s$ configuration, i.e. $c_2=0$ and $c_2=\frac{1}{2}$ CFTs as stated above. Firstly we consider the BTW model on the Ising-correlated percolation lattice. This produces Ising-like correlations in the diluteness pattern of the lattice which is controlled by an artificial temperature $T$. To build this lattice we consider the ferromagnetic Ising model in the zero magnetic field $H=-\sum_{\left\langle i,j\right\rangle }\sigma_i\sigma_j$ (the coupling constant has been set to unity) in which $\left\langle i,j\right\rangle$ means that the sites $i$ and $j$ are neighbors and $\sigma_i\equiv 2(s_i-1/2)$ is the artificial spin taking two values: $\sigma=1$ for the active sites (corresponding to $s=1$) and $\sigma=-1$ for the inactive sites (corresponding to $s=0$). In this case the correlations between $s$-fields generate some non-trivial effects. It is well-known that the Ising model shows two simultaneous transitions at some critical temperature $T_c$; the magnetic and the percolation transitions. The percolation transition occurs for the geometrical connected spin clusters which are composed of spins with the same signs ($\sigma=1$ for our case) which are connected~\cite{Najafi2016Monte}. At $T=T_c$ a percolated cluster can surely be found. At this temperature the 2D Ising model is described by $c=\frac{1}{2}$ conformal field theory which belongs to the minimal conformal series with $p=3$. In addition, we simulate the BTW model on the uncorrelated critical site percolation theory ($c_2=0$); namely we set a typical site active ($s=1$) with the probability $p_c$ and inactive ($s=0$) with the probability $1-p_c$, in which $p_c$, in which $p_c$ is the critical occupation of the percolation (percolation threshold) at which a (second order) transition occurs, i.e. a connected cluster of $s=1$ state which percolates throughout the sample can surely be found. In this case the diluteness pattern is uncorrelated. \\
The aim of the present paper is to seek for the emergent fixed points to which the $-2\oplus \frac{1}{2}$ (BTW on the Ising-correlated site-diluted percolation lattice) and $-2\oplus 0$ (BTW on the uncorrelated site-diluted percolation lattice) correspond.

\section{Numerical tests and SLE}\label{num}
\label{sec:num}

\begin{figure}
	\centerline{\includegraphics[scale=.4]{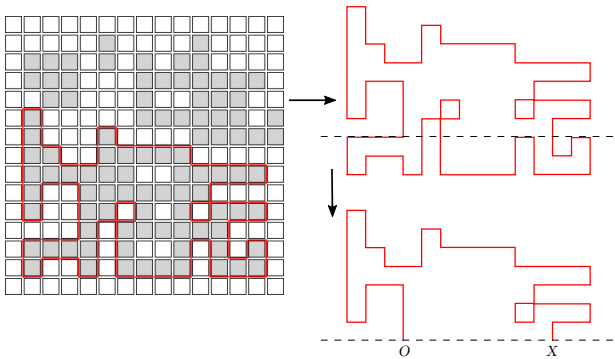}}
	\caption{The cutting procedure of a loop to obtain an open stochastic curve.}
	\label{fig:cutting}
\end{figure}

Our analysis in this paper is restricted to the external frontiers of the avalanches (on both lattices) which form conformal loop ensemble (CLE). These loops are simply the separators of the toppled and un-toppled sites in an avalanche. This analysis which is proved to be fruitful, reflects the internal symmetries of the model in hand. The most important quantity is the fractal dimension ($D_f$) of the loops. To define this quantity let us represent the loop lengths by $l$ and the gyration radius of the loops by $r$. The fractal dimension is defined via the relation $\left\langle \log(l)\right\rangle =D_f\left\langle \log(r)\right\rangle$. The other important quantity is the loop Green function $G|i-j|$ which is defined as the probability that the sites $i$ and $j$ belong to the same loop. Both of these quantities are analyzed in the subsequent sections. In some parts of the paper, we need the open random curves which start from the origin and end on some point at infinity. There is well-known strategy to obtain such curves from the loops. According to this idea one cuts the loop, e.g. from its center of mass in some arbitrary direction. Due to the rotational invariance of the system, it is not important what this direction is. After this process we pick a portion of the loop and embed it in the upper half plane in such a way that the resulting curve starts from the origin and ends on the real axis, namely $X$. This procedure has been shown schematically in Fig. \ref{fig:cutting} in which the gray area is the set of toppled sites whose boundary has been shown by a red loop. The resulting open curve has been extended from the origin to the real point $X$. By applying the map $w(z)=X\frac{z}{z-X}$, ($z=u+iv$ is the coordinate of the upper-half plane) we will have the desired curve. Note that this map preserves the origin, and sends the end point $z=X$ to the infinity. Having the open curves, one can parameterize the curve by \textit{time} $t$ and calculate the $\nu$ exponent which is defined by the relation $\left\langle R^2\right\rangle \propto t^{2\nu}$ in which $R$ is the distance of the curve at time $t$ from the origin. Also LPP and WA as well as direct SLE mapping can be used to extract the diffusivity parameter $\kappa$ of the SLE theory. According to the SLE theory one can describe the geometrical objects (which may be interfaces) of a 2D critical model via a growth process and classify them into one parameter ($\kappa$) classes \cite{Cardy}. From a simple relation between the central charge $c$ in CFT and the diffusivity parameter $\kappa$ in SLE, namely $c=\frac{(6-\kappa)(3\kappa-8)}{2\kappa}$, one can find the corresponding CFT \cite{Cardy,NajafiPRE1,NajafiPRE2}, and consequently the universality class is obtained. Chordal SLE$_{\kappa}$ is a growth process defined via the conformal maps, $g_{t}(z)$, which are solutions of the Loewner's equation $\partial_{t}g_{t}(z)=\frac{2}{g_{t}(z)-\xi_{t}}$ where the initial condition is $g_{t}(z)=z$ and $\xi_{t}$ (the driving function) is a continuous real valued function which is shown to be proportional to the one dimensional Brownian motion ($\xi_t=\sqrt{\kappa}B_t$) if the curves have two properties: conformal invariance and the domain Markov property. SLE($\kappa,\kappa-6$) is a variant of SLE$_{\kappa}$ for which the curves go from the origin to a point on the real axis $X$ and the driving function satisfies the relation $\text{d}\xi_t=\sqrt{\kappa}\text{d}B_t+\frac{\kappa-6}{\xi_t-g_t(X)}\text{d}t$. This method has been shown to by more precise in the numerical analysis, for which the best amount of $\kappa$ is chosen so that $B_t$ in this equation is fitted as good as possible to the one-dimensional Brownian motion, see~\cite{Najafi2012observation,NajafiPRE1}. The LPP~\cite{NajafiPRE1,NajafiPRE2} and the WA~\cite{Boffetta} statistics are some other important tests for extracting $\kappa$. In the former case the probability that a point $z_0=u_0+iv_0$ falls to the right of a random open curve which is extended from the origin to the infinity is calculated. It is shown that LPP$(\theta)=\frac{1}{2}+\frac{\Gamma(4/\kappa)}{\sqrt{\pi}\Gamma((8-\kappa)/2\kappa)}{_2}F_1\left[\frac{1}{2},\frac{4}{\kappa},\frac{3}{2},-\left( \cot(\theta)\right)^2 \right] \cot(\theta)$~\cite{NajafiPRE1} in which $\cot(\theta)\equiv \frac{u_0}{v_0}$. For the WA statistics, let us take $\theta$ as the angel between the vector which is extended from the origin to a point of the open curve ($\vec{r}$), and the local tangent vector on the curve. It is shown that $\left\langle\theta^2 \right\rangle=a+\frac{2\kappa}{8+\kappa}\log(r)$ and $\left\langle\theta \right\rangle=0$~\cite{Boffetta}. The fractal dimension can also be interpreted as another measure which is related to $\kappa$ via the relation $D_F=1+\frac{\kappa}{8}$~\cite{Cardy}. In the following sections we apply these tests to extract $\kappa$ for both models.

\section{Summation of $c=-2$ and $c=1/2$ CFTs}\label{sum1}

\begin{figure*}
	\centering
	\begin{subfigure}{0.45\textwidth}\includegraphics[width=\textwidth]{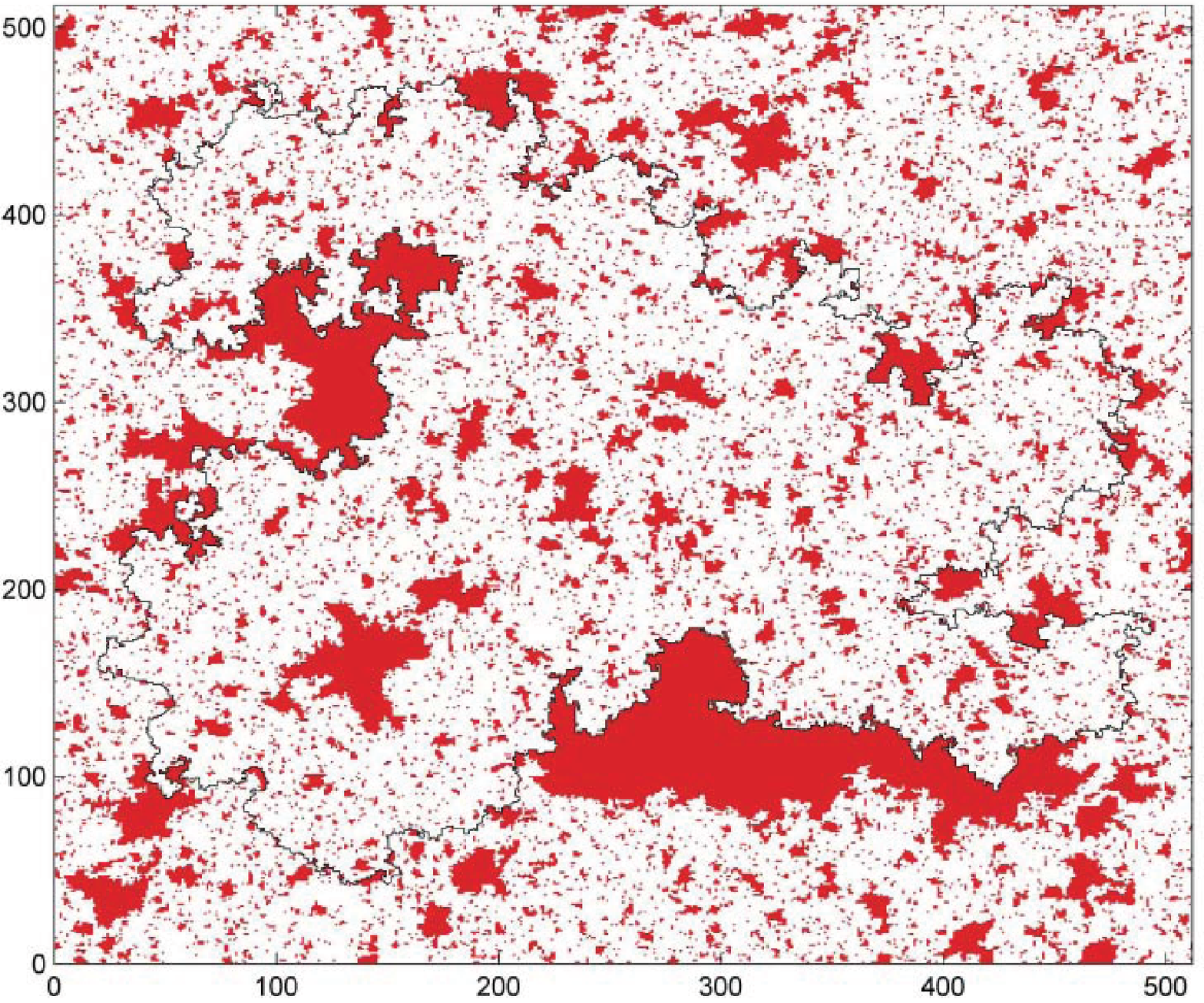}
		\caption{}
		\label{fig:512Tc}
	\end{subfigure}
	\centering
	\begin{subfigure}{0.45\textwidth}\includegraphics[width=\textwidth]{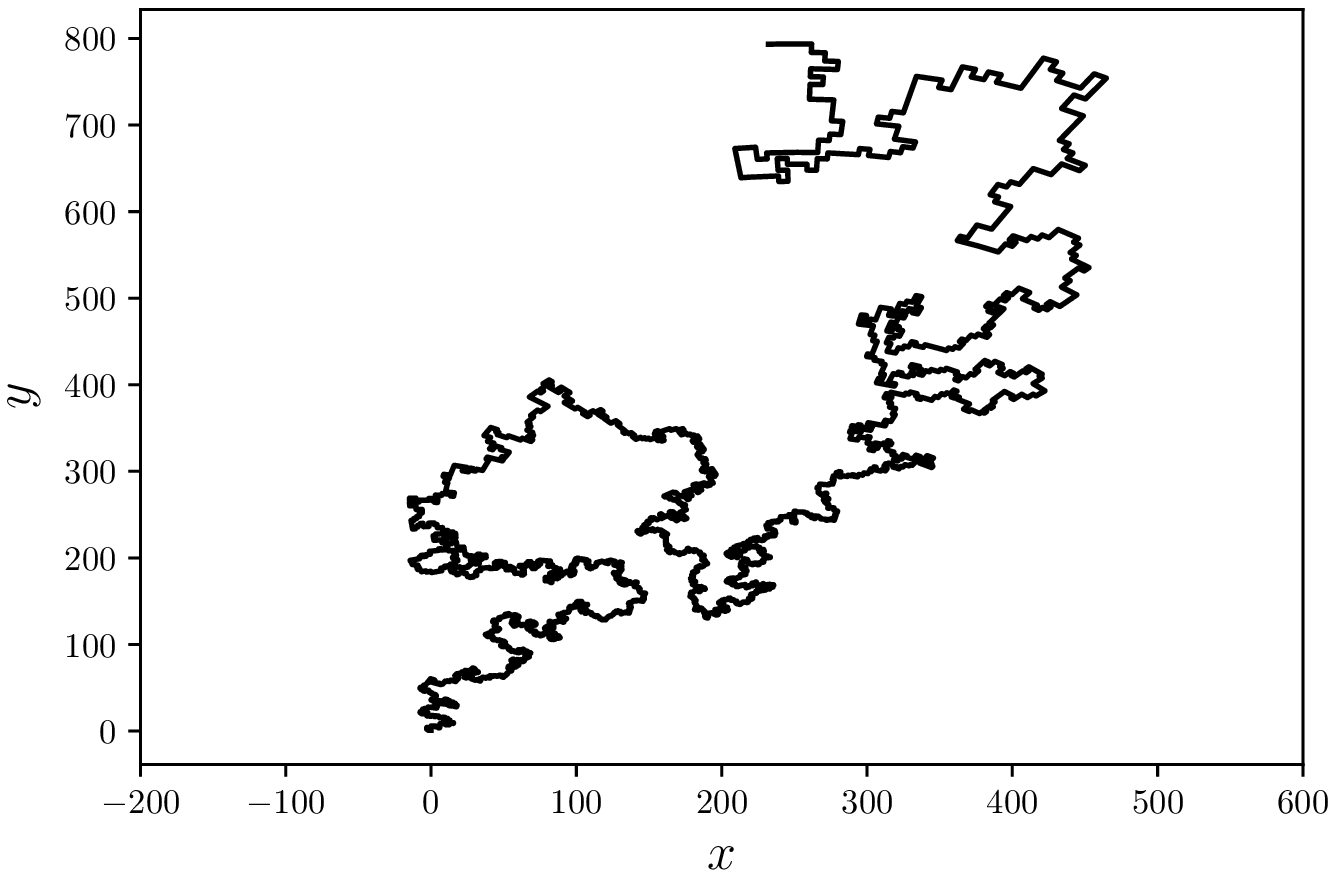}
		\caption{}
		\label{fig:mapped-sample}
	\end{subfigure}
	\caption{(Color online): (a) A $500\times 500$ sample of the BTW on the Ising-percolated cluster. The red-sites are spin-down sates and the white sites are spin-up states. (b) A sample of the mapped cut stochastic curves.}
	\label{fig:samples}
\end{figure*}

In this section we simulate the BTW automaton model in which the particles are allowed to move only through the quenched \textit{up-spin} sites of the critical square Ising lattice. These spins have been obtained using the Ising model at the critical artificial temperature $T=T_c$ which is nearly $2.269$ \cite{Goldenfeld} for the square lattice. We note that there is no real magnetic sites, instead the spins show activity of the sites and the correlations in the diluteness pattern of sites are controlled by the Ising model. For the simulations we have used the Wolff Monte Carlo algorithm to avoid the critical slowing down. To make Ising samples uncorrelated, we have generated $\frac{L}{3}$ random spin flips between two successive samplings. The samples have been generated in square lattices with linear sizes $L=128, 256, 512, 1024$ and $2048$ and for each lattice size $5\times 10^6$ avalanches have been generated with $100$ uncorrelated Ising samples. Therefore our ensemble averages involve averaging over both BTW avalanches and Ising samples. A typical sample of BTW on a critical Ising lattice has been shown in Fig. \ref{fig:512Tc} in which the white sites are in the \textit{up-spin} (active) state and the red ones show the \textit{down-spin} (inactive) sites through which the sand grains cannot pass. The toppled region in a single avalanche is distinguished from the un-toppled ones via a black loop in this figure which shows the external frontier of the avalanche. We have extracted and analyzed these loops which are expected to determine its presumable universality class. A stochastic sample of mapped (via the mapping $w(z)$) cut curve which is open has been shown in \ref{fig:mapped-sample} in a $800\times 800$ box. To avoid the numerical errors we have considered these curves up to the point in which the deformed lattice parameter exceeds $10$ times the regular lattice parameter. The fitting parameters have been obtained by means of the $\chi^2$ method.

\begin{figure*}
	\centering
	\begin{subfigure}{0.45\textwidth}\includegraphics[width=\textwidth]{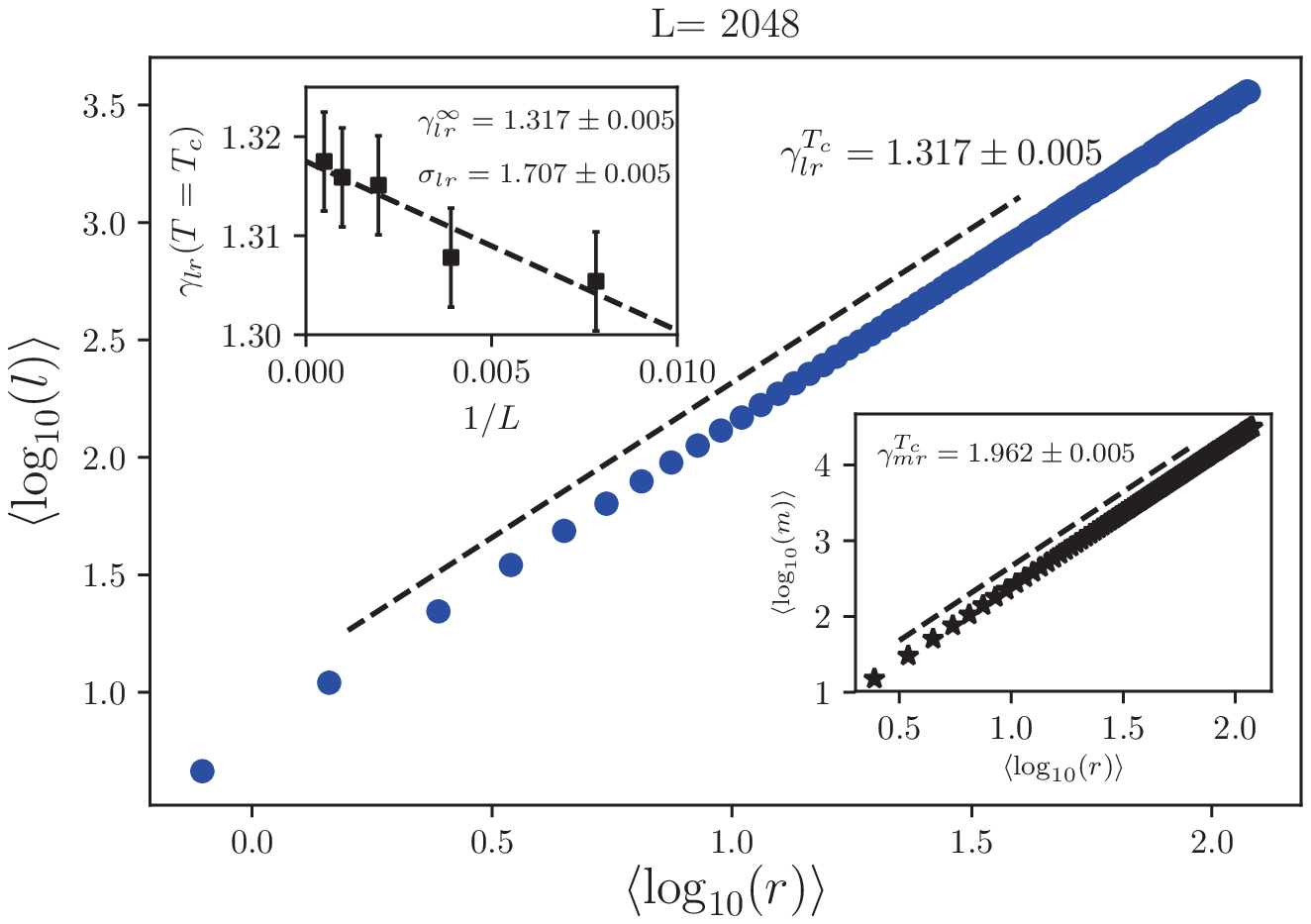}
		\caption{}
		\label{fig:Df}
	\end{subfigure}
	\centering
	\begin{subfigure}{0.45\textwidth}\includegraphics[width=\textwidth]{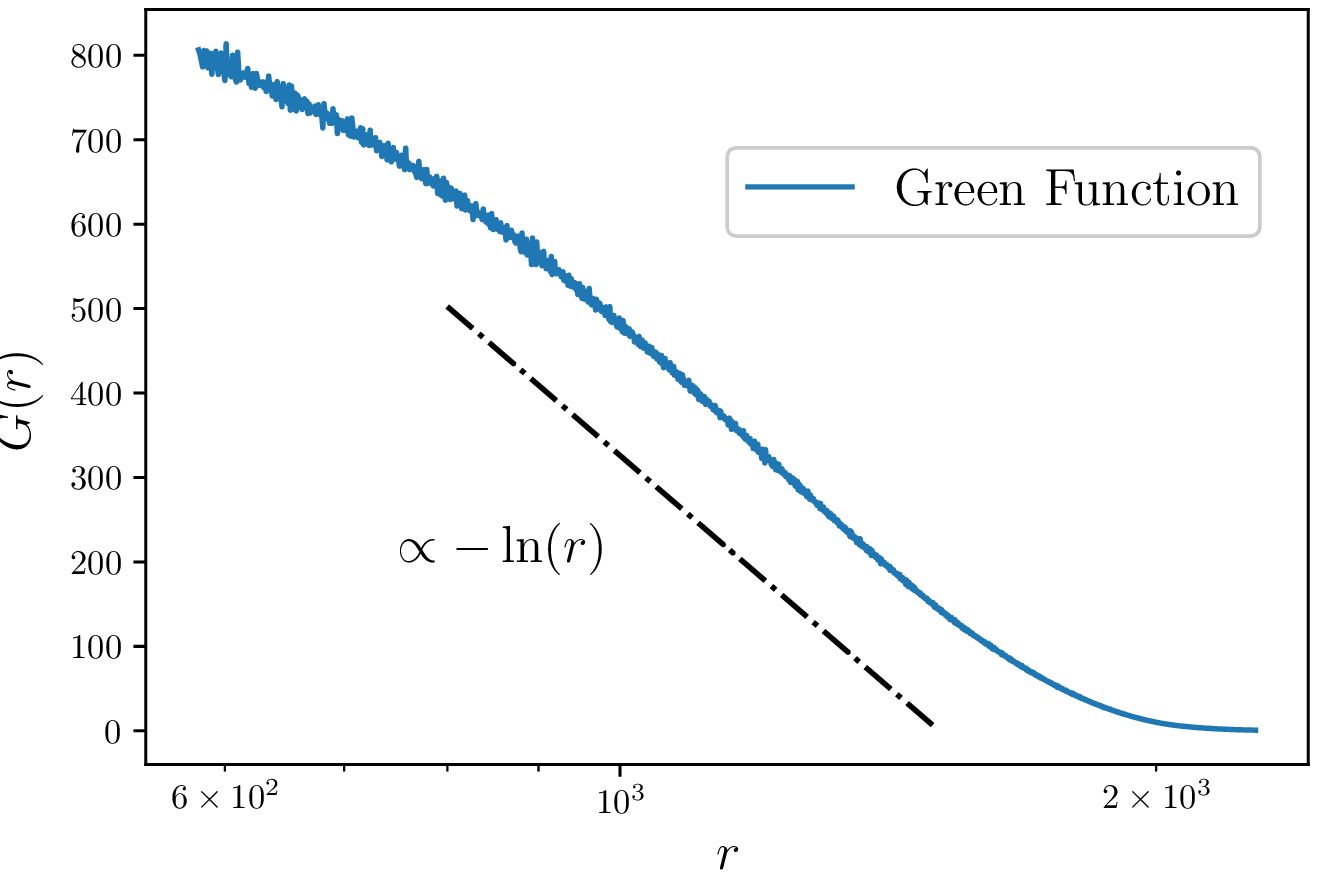}
		\caption{}
		\label{fig:Green}
	\end{subfigure}
	\centering
	\begin{subfigure}{0.45\textwidth}\includegraphics[width=\textwidth]{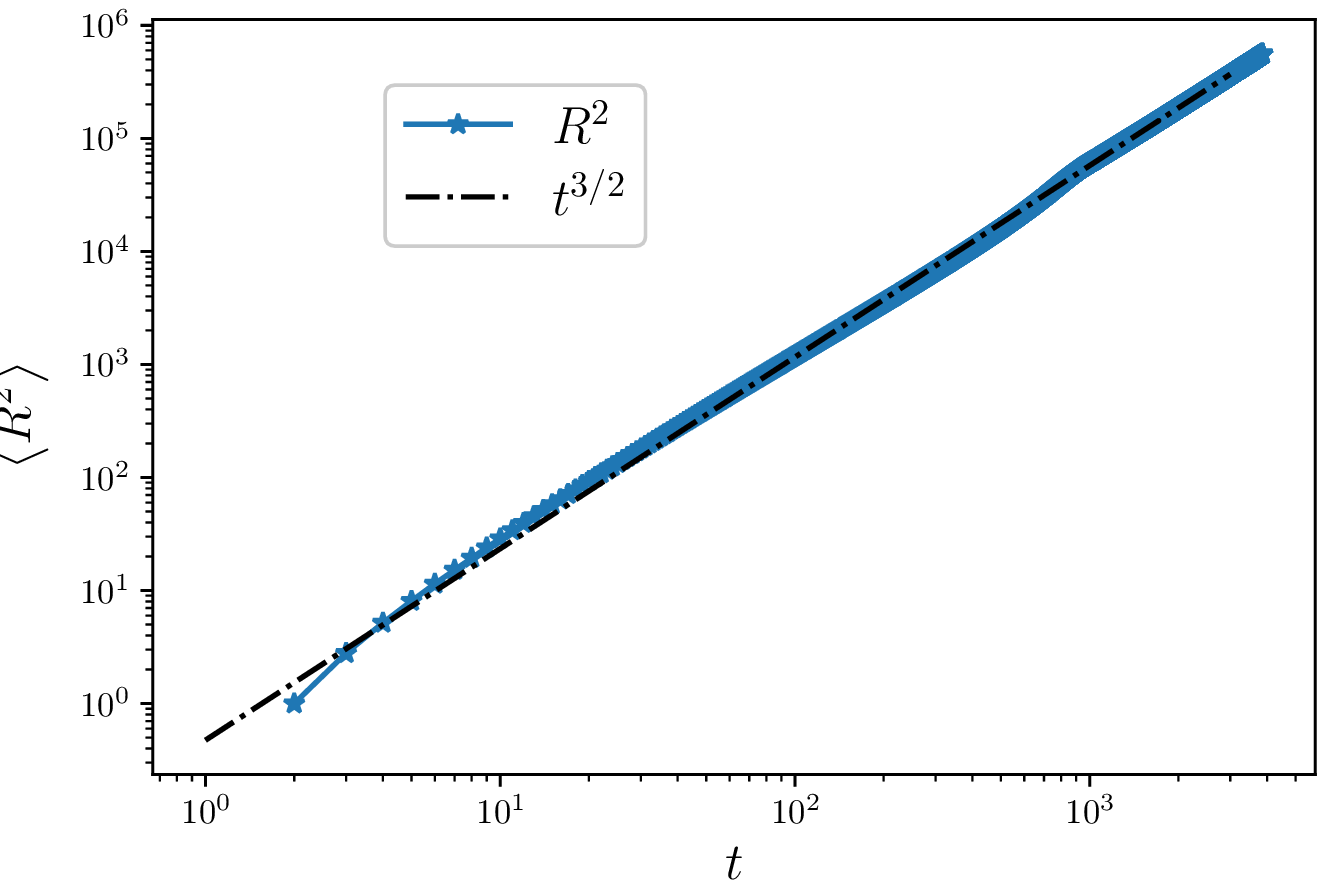}
		\caption{}
		\label{fig:R2}
	\end{subfigure}
	\centering
	\begin{subfigure}{0.45\textwidth}\includegraphics[width=\textwidth]{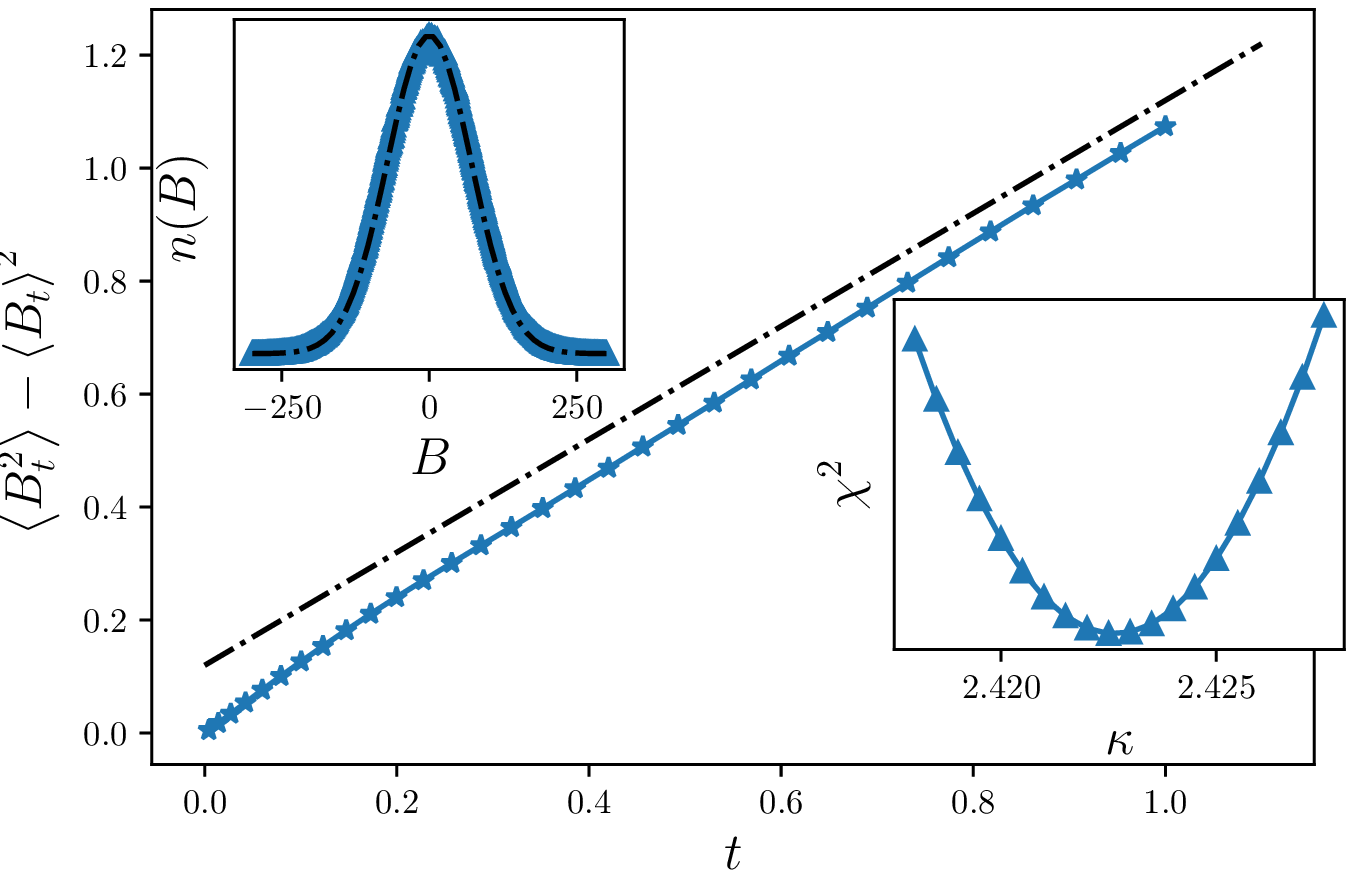}
		\caption{}
		\label{fig:SLEkr}
	\end{subfigure}
	\centering
	\begin{subfigure}{0.45\textwidth}\includegraphics[width=\textwidth]{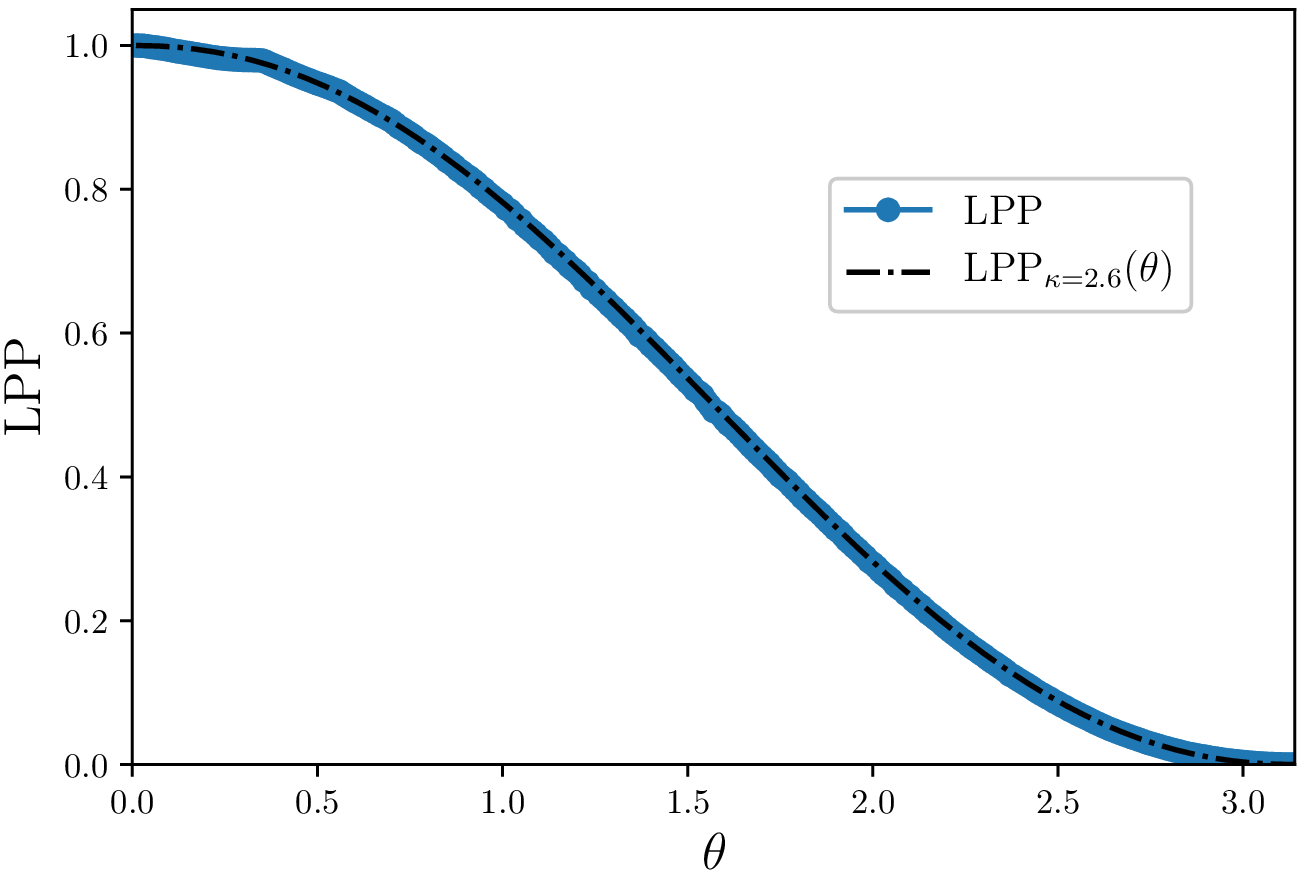}
		\caption{}
		\label{fig:LPP}
	\end{subfigure}
	\centering
	\begin{subfigure}{0.45\textwidth}\includegraphics[width=\textwidth]{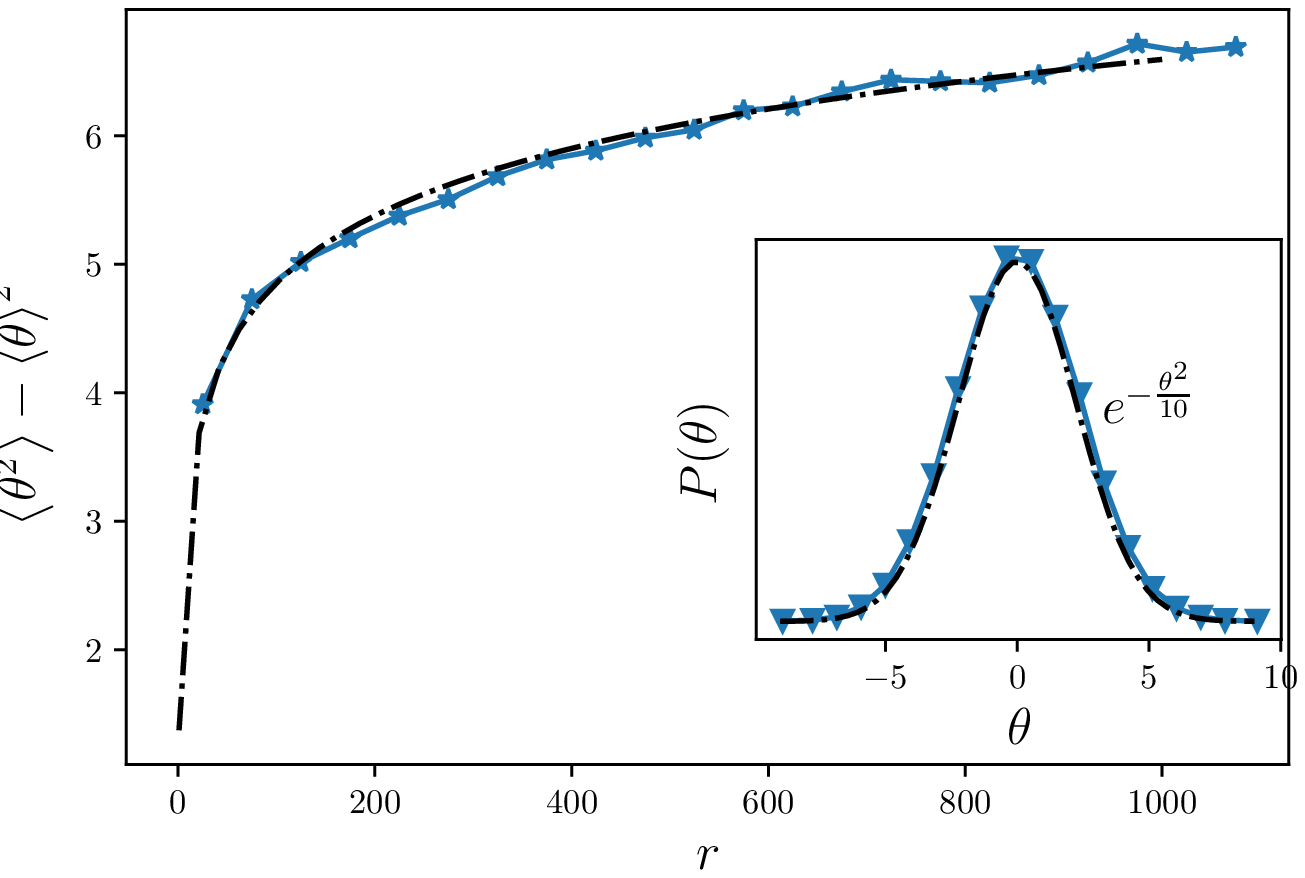}
		\caption{}
		\label{fig:WA}
	\end{subfigure}
	\caption{(Color online): (a) Fractal dimension for the BTW model on the Ising-percolation lattice at $T=T_C$. Up inset: Finite size dependence of $\gamma_{lr}\equiv D_F$ with its asymptotic values: $\gamma_{lr}(L)=\gamma_{lr}^{\infty}-\sigma_{lr}(1/L)$. Lower inset: The $\gamma_{mr}$ exponent. (b) The Green function $G(r)$ in terms of $r$. (c) $\left\langle R^2\right\rangle $ in terms of $t$ for open curves, with the exponent $\nu=3/4$. (d) The variance of $B_t$ for $\kappa=2.42$. Upper inset: the distribution function $n(B)$ for $t=5000$. Lower inset: the amount of $\chi^2$ in terms of $\kappa$ with a minimum at $\kappa=2.42$. (e) LPP in terms of $\theta$ and its fit by $\kappa=2.6$. (f) The variance of WA in terms of $r$. Inset: The distribution of $\theta$, i.e. $P(\theta)$ for $r=250$ which is compatible with the Gaussian distribution function $n(\theta)\propto \exp\left[\theta^2/(2\sigma)\right]$ with $\sigma=a+\frac{2\kappa}{8+\kappa}\log(r)$.}
	\label{fig:stat}
\end{figure*}

The results for the fractal dimension $\gamma_{lr}\equiv D_F$ along with the finite size effects have been appeared in Fig. \ref{fig:Df}. In the upper inset we have shown the $1/L$-dependence of this exponent whose linear dependence yields $\gamma_{lr}(L\rightarrow\infty)\equiv D_f=1.317\pm 0.005$. In the lower inset we have shown the power-law dependence of the mass of the avalanches $m$ (defined as the total number of toppled sites in an avalanche) to $r$ with the exponent $\gamma_{mr}^{T_c}=1.962\pm 0.005$. In Fig.\ref{fig:Green} the Green function has been shown in terms of $r$ for $L=2048$. We see that for intermediate values of $r$, the dependence is logarithmic, in accordance with ordinary BTW model. The closest fractional value of the obtained $D_F$ is $\frac{4}{3}$ which is the fractal dimension of the 2D self-avoiding walk (SAW). To test it more directly we have calculated the exponent $\nu$ for the open mapped cut curves which is perfectly compatible with the numerical value $\nu_{\text{SAW}}=\frac{3}{4}$ as is evident in the Fig.~\ref{fig:R2}. Note that by using some simple scaling arguments one finds that $D_f=\frac{1}{\nu}$. \\
In the remaining of this section we address the conformal invariance of the resultant model. Firstly we have used the direct SLE($\kappa,\kappa-6$) method with the slit mapping. For the details of the application of SLE($\kappa,\kappa-6$) to the statistical models see~\cite{Najafi2012observation}. To test the $B_t$ function (which has been defined in the previous section and is supposed to be properly fitted by a Brownian motion) we have used the $\chi^2$ method and the results have been shown in Fig.~\ref{fig:SLEkr}. From the lower inset it is evident that $\kappa=2.4\pm 0.1$ which is compatible with $\kappa_{\text{SAW}}=\frac{8}{3}$. In the upper inset the distribution of $B$, i.e. $n(B)$ has been sketched for $t=5000$ which is precisely compatible with the one-dimensional Brownian motion whose form is the Gaussian. \\
The other important test is LPP which has been presented in Fig.\ref{fig:LPP}. The analytical exact solution has been shown for $\kappa=2.6$, demonstrating that the best fit is obtained for the SAW. The WA test has also been done for which the results have been gathered in Fig.~\ref{fig:WA}. The best fit has been obtained for the analytical relation with $\kappa=2.5$ which shows again that SAW is the closest model for the $-2\oplus \frac{1}{2}$ fixed point. The results have been gathered in TABLE~\ref{tab:4d-tau}.

\begin{table}
	\caption{The numerical amounts of $\kappa$ which have been obtained by various SLE tests. The exact result for SAW has been shown for comparison.}
	\label{tab:4d-tau}
	\begin{center}
		\begin{tabular}{|c|c|c|c|c|c|}
			\hline & frac. dim. & direct & LPP & WA & 2D SAW\\
			\hline $\kappa$ & $2.56(4)$ & $2.4(1)$ & $2.6(1)$ & $2.5(2)$ & $\frac{8}{3}$\\
			\hline
		\end{tabular}
	\end{center}
\end{table}

\section{Summation of $c=-2$ and $c=0$ CFTs}\label{sum2}
\begin{figure*}
	\centering
	\begin{subfigure}{0.45\textwidth}\includegraphics[width=\textwidth]{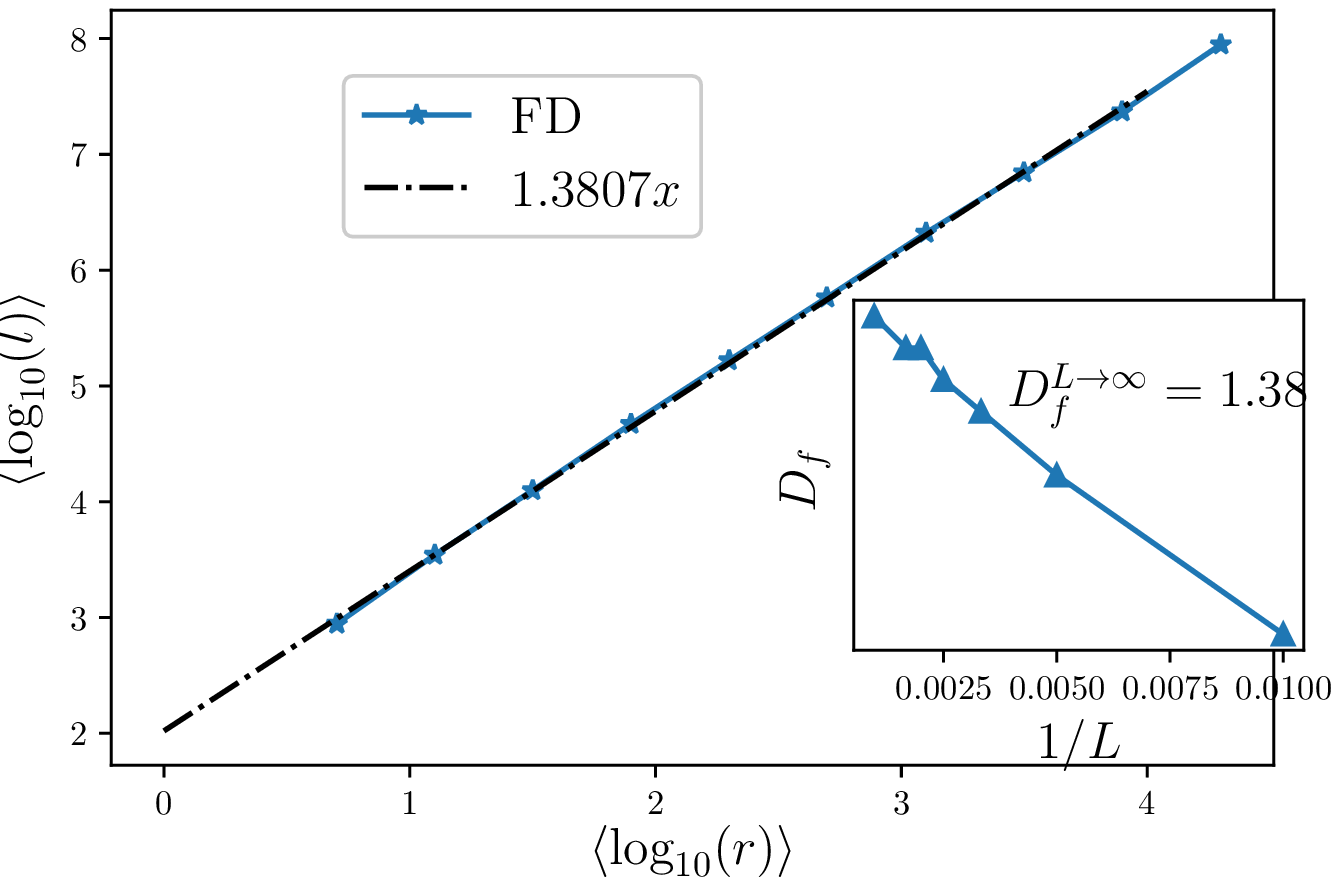}
		\caption{}
		\label{fig:FD}
	\end{subfigure}
	\centering
	\begin{subfigure}{0.45\textwidth}\includegraphics[width=\textwidth]{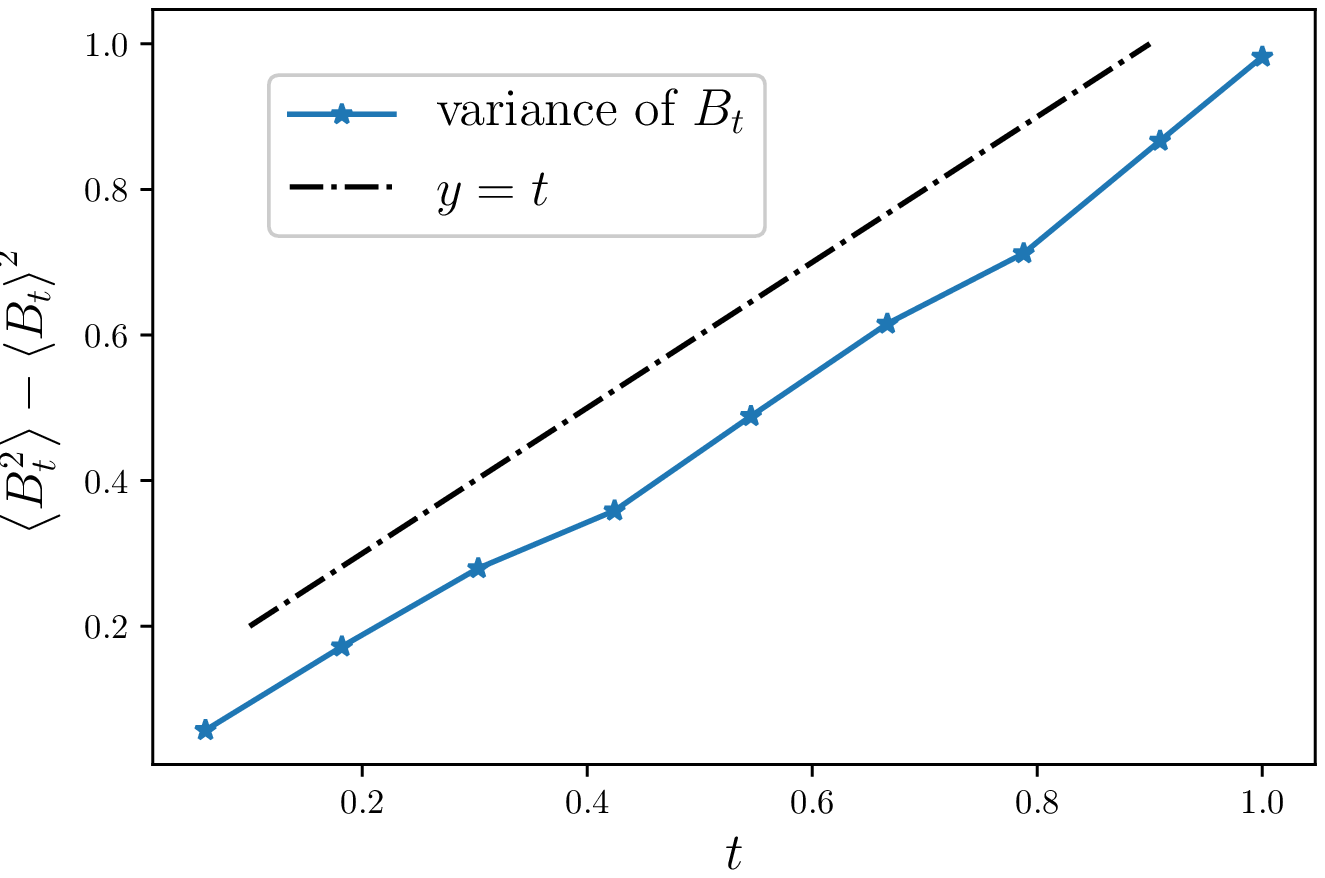}
		\caption{}
		\label{fig:SLE}
	\end{subfigure}
	\caption{(Color online): (a) The fractal dimension of the exterior frontiers of the BTW model on the uncorrelated percolation lattice. Inset: the finite size dependence. (b) The variance of $B_t$ for $\kappa=3.0$.}
	\label{fig:c=0}
\end{figure*}
Let us now consider the coupling of the BTW mode and the critical site percolation theory, i.e.  $-2\oplus 0$. The percolation lattice is defined as the square lattice with the critical occupation parameter $p=p_c$. Such an investigation has been previously done in \cite{Najafi2016Bak,Najafi2013water}, supporting the hypothesis that the resultant fixed point is compatible with the Ising universality class. Since the full treatment has been done in \cite{Najafi2013water}, we have sufficed to two SLE tests in this section: the fractal dimension and the direct SLE mapping methods. In the Fig.~\ref{fig:FD} we have shown the results for the fractal dimension $D_F$. In the inset of this figure the finite-size dependence of this quantity has been shown from which we see that $D_F^{L\rightarrow\infty}=1.38$ whose closest fractional value is $\frac{11}{8}$ which is the fractal dimension of the external perimeter of the geometrical spin clusters of the critical Ising model. It is well-known that for the critical Ising model $\kappa=3$ which has been tested directly by slit mapping of the SLE($\kappa,\kappa-6$) method. We have found that $B_t$ is best fitted to the Brownian motion for $\kappa=3.0\pm 0.3$ which has been shown in Fig.~\ref{fig:SLE}. For the full treatment see~\cite{Najafi2013water}.\\

\section*{Summary and Discussion}
\label{sec:conc}

In this paper we have considered numerically the coupling of $-2\oplus \frac{1}{2}$ and $-2\oplus 0$ CFTs. We have considered the 2D BTW as a prototype of $c=-2$ LCFT, the uncorrelated site percolation as a prototype of $c=0$ LCFT and the critical Ising model as a prototype of $c=\frac{1}{2}$ CFT. The Schramm-Loewner evolution (SLE) theory has been employed to address the conformal invariance of these models. Using precise SLE tests, namely direct SLE mapping, fractal dimension, left passage probability (LPP) and winding angel (WA) tests we have extracted the diffusivity parameter $\kappa$ of the fixed point $-2\oplus \frac{1}{2}$. We found that the results are compatible with the self-avoiding walk (SAW) fixed point, for which $\kappa=\frac{8}{3}$ and $c=\tilde{0}$. We have used the symbol $\tilde{0}$, since $\kappa_{\text{SAW}}=\frac{8}{3}$ is related to the $\kappa_{\text{percolation}}=6$ with the duality relation $\kappa_{\text{SAW}}=\frac{16}{\kappa_{\text{percolation}}}$ both of which correspond to the same CFT class, namely $c=0$.\\
The results of $-2\oplus 0$~\cite{Najafi2013water} have been produced in this paper to be self-contained. The resulting fixed point is compatible with the $c=\frac{1}{2}$ CFT class. These results can be gathered in the following relations, which show the closeness of the central charges by the operation of couplings:
\begin{equation}
\begin{split}
-2\oplus 0=\frac{1}{2}\\
-2\oplus \frac{1}{2}=\tilde{0}.
\end{split}
\end{equation}
It is notable that the Ising model has been simulated on the uncorrelated percolation lattice  in the Ref. \cite{Najafi2016Monte}. However no highlighted result has been reported for  $p=p_c$ to find the coupling $\frac{1}{2}\otimes 0$ and the problem is open.\\
\acknowledgments
We thank J. Cheraghalizadeh for helping in some simulations in the paper.

\end{document}